# Reconnection-Driven Coronal-Hole Jets with Gravity and Solar Wind

J. T. Karpen[1], C. R. DeVore[1], S. K. Antiochos[1], & E. Pariat[2]

[1]Heliophysics Science Division, NASA Goddard Space Flight Center, Greenbelt MD 20771

[2] LESIA, Observatoire de Paris, PSL Research University, CNRS, Sorbonne Université, UPMC Univ. Paris 06, Univ. Paris Diderot, Sorbonne Paris Cité, 5 place Jules Janssen, 92195 Meudon, France

**Abstract**

Coronal-hole jets occur ubiquitously in the Sun's coronal holes, at EUV and X-ray bright points associated with intrusions of minority magnetic polarity. The embedded-bipole model for these jets posits that they are driven by explosive, fast reconnection between the stressed closed field of the embedded bipole and the open field of the surrounding coronal hole. Previous numerical studies in Cartesian geometry, assuming uniform ambient magnetic field and plasma while neglecting gravity and solar wind, demonstrated that the model is robust and can produce jet-like events in simple configurations. We have extended these investigations by including spherical geometry, gravity, and solar wind in a nonuniform, coronal hole-like ambient atmosphere. Our simulations confirm that the jet is initiated by the onset of a kink-like instability of the internal closed field, which induces a burst of reconnection between the closed and external open field, launching a helical jet. Our new results demonstrate that the jet propagation is sustained through the outer corona, in the form of a traveling nonlinear Alfvén wave front trailed by slower-moving plasma density enhancements that are compressed and accelerated by the wave. This finding agrees well with observations of white-light coronal-hole jets, and can explain microstreams and torsional Alfvén waves detected in situ in the solar wind. We also use our numerical results to deduce scaling relationships between properties of the coronal source region and the characteristics of the resulting jet, which can be tested against observations.

## I. Introduction

Jets are observed to occur frequently in all regions of the solar atmosphere, from coronal holes to quiet Sun to active regions. These events are most readily observed in coronal holes, where they frequently are referred to as coronal-hole jets, due to the contrast between their enhanced EUV and X-ray emissions and the low background provided by the surrounding cool, tenuous plasma. Every spaceborne solar observatory with UV, EUV, or X-ray imaging and/or spectroscopy has observed jets, from *Skylab* (Schmahl 1981) to *HRTS* (Dere et al. 1989, 1991), *Yohkoh* (Shibata et al. 1992, 1994; Shimojo et al. 1996, 1998), *SOHO* (Wang et al. 1998; Wood et al. 1999; Wang & Sheeley 2002), *TRACE* (Chae et al. 1999), *STEREO* (Patsourakos et al. 2008; Nisticò et al. 2009, 2010; Feng et al. 2012), *Hinode* (Cirtain et al. 2007; Savcheva et al. 2007; Liu et al. 2009; Török et al. 2009; Doschek et al. 2010; Moore et al. 2010; Liu et al. 2011; Moore et al. 2013; Sterling et al. 2015), *SMEI* (Yu et al. 2014), *SDO* (Hong et al. 2011, 2013; Young & Muglach 2014), and *IRIS* (Cheung et al. 2015). A broad survey of observations, theory, and modeling of solar jets is available (Raouafi et al. 2016).

Many EUV and/or X-ray jets, particularly larger-scale events occurring close to the solar limb, also exhibit signatures observed by white-light coronagraphs (Wang et al. 1998; Wood et al. 1999; Wang & Sheeley 2002; Patsourakos et al. 2008; Nisticò et al. 2009, 2010; Hong et al. 2011; Feng et al. 2012; Hong et al. 2013; Yu et al. 2014). These observations show conclusively that coronal-hole jets have impacts well beyond their point of origin in the inner corona. Indeed, they may be sufficiently numerous to provide a significant fraction of the total mass, momentum, and energy carried away by the solar wind (Moore et al. 2011). The individual, small-scale disturbances of the solar wind referred to as microstreams (Neugebauer et al. 1995) may be interplanetary signatures of coronal jets (Neugebauer 2012). Furthermore, because many coronal-hole jets have been observed to exhibit helical motions consistent with nonlinear waves (Patsourakos et al. 2008; Liu et al. 2009; Nisticò et al. 2009), they could be the source of torsional Alfvén waves measured in situ in the interplanetary medium (Gosling et al. 2010; Marubashi et al. 2010).

In previous work (Pariat et al. 2009, 2010, 2015, 2016; hereafter P09, P10, P15, and P16, respectively; Dalmasse et al. 2012), we proposed the embedded-bipole model for coronal-hole jets, in which the event is triggered by the sudden onset of explosive reconnection in a quasi-equilibrium magnetic configuration. The source region consists of a small, relatively strong concentration of minority-polarity vertical flux embedded in a surrounding broad sea of more diffuse majority-polarity flux. Such structures are sprinkled liberally throughout coronal holes. The plasma exhibits the morphology of bright tendrils emanating from a central locus: these tendrils illuminate magnetic loops that depart from the concentrated minority flux and close back to the Sun nearby in the diffuse majority flux. The loops form a dome-shaped structure of closed magnetic field with a null point on its surface. Beyond the dome, the field lines extend into the outer corona sufficiently far that they are opened to the heliosphere by the solar wind. This separation of the source structure into two distinct flux systems, one open and one closed, enables electric currents to be established readily by relative displacements of the field inside and outside of the dome (Antiochos 1990; Lau & Finn 1990; Antiochos 1996). When sufficiently strong, these currents initiate magnetic reconnection, accompanied by mass motions and plasma heating. If the reconnection is explosive, a nonlinear helical Alfvén wave ensues, accompanied by upflows of dense plasma. Throughout this paper, we use the term "jet" to describe the entire dynamic event, encompassing both the leading Alfvén wave and the slower, dense outflow.

These investigations to date have demonstrated the essential viability of the embedded-bipole model for coronal-hole jets. However, we made several simplifying assumptions: Cartesian geometry, no gravity, a static initial atmosphere, and initially uniform background magnetic field and plasma density. The neglect of gravity is especially important, because it implies that the jet speeds at large heights may well be overestimated. In order to test the embedded-bipole model rigorously and to develop a robust understanding of jet evolution and propagation in the outer corona and inner heliosphere, more realistic numerical simulations must be performed. In our present study, therefore, we incorporate the effects of spherical geometry, gravitational acceleration, density stratification, and ambient magnetic field variation on the reconnection-driven jet dynamics in the embedded-bipole model. In addition, we exploit the advantage of dynamic adaptive mesh refinement in these simulations. As is demonstrated in detail below, the model proves to be robust and produces vigorous, fast jets in this much more realistic, global-scale setting.

We describe the physical and numerical model in §2. The results of the numerical simulations are presented in §3. Analysis of our results and their scaling to solar observations of jets are given in §4. We discuss the implications of our findings and prospects for further research in §5.

**II. Model**

We performed three-dimensional (3D) numerical simulations of reconnection-driven solar coronal-hole jets using the Adaptively Refined Magnetohydrodynamics Solver (DeVore & Antiochos 2008), which we used in our previous Cartesian studies (see §1). ARMS employs the widely used PARAMESH toolkit (MacNeice et al. 2000; Olson & MacNeice 2005) to manage the parallel processing and adaptive meshing aspects of the solver, and well-established Flux Corrected Transport algorithms (DeVore 1991) to advance the solutions in time. For this investigation, we solved the time-dependent equations of ideal, compressible, isothermal MHD in the form

$$\frac{\partial \rho}{\partial t} + \nabla \cdot (\rho \mathbf{v}) = 0,$$

$$\frac{\partial \rho \mathbf{v}}{\partial t} + \nabla \cdot (\rho \mathbf{v}\mathbf{v}) = -\nabla P + \rho \mathbf{g} + \frac{1}{4\pi}(\nabla \times \mathbf{B}) \times \mathbf{B}, \quad (1)$$

$$\frac{\partial \mathbf{B}}{\partial t} - \nabla \times (\mathbf{v} \times \mathbf{B}) = 0,$$

with time $t$, mass density $\rho$, velocity $\mathbf{v}$, thermal pressure $P$, gravitational acceleration $\mathbf{g}$, and magnetic field $\mathbf{B}$. The solar gravity is given by

$$\mathbf{g} = -g_e \frac{R_e^2}{r^2} \hat{r}, \quad (2)$$

where $g_\odot = 2.7 \times 10^4$ cm s$^{-2}$ is the value at radius $R_\odot = 7 \times 10^{10}$ cm, and $r$ is the spherical radial coordinate. The ideal-gas law for fully ionized hydrogen is

$$P = 2 \frac{k_B}{m_p} \rho T, \quad (3)$$

where $k_B$ and $m_p$ are the Boltzmann constant and proton mass, respectively. The assumption of constant and uniform temperature $T$ then closes the system of equations.

The isothermal approximation yields the simplest representation of the solar wind. However, it also neglects heating associated with the reconnection-driven outflows that comprise our jet. Therefore, we cannot definitively predict observables that depend upon the detailed thermodynamical behavior of the coronal plasma, such as its optically thin radiative signatures. As in other work (Masson et al. 2013), we initialized the atmosphere with the steady, spherically symmetric, supersonic wind (Parker 1958)

$$\frac{v_r^2}{c_s^2} \exp\left(1 - \frac{v_r^2}{c_s^2}\right) = \frac{r_s^4}{r^4} \exp\left(4 - 4\frac{r_s}{r}\right). \quad (4)$$

The radial velocity is $v_r$, and the isothermal sound speed $c_s$ and sonic point $r_s$ are given by

$$c_s^2 = 2\frac{k_B}{m_p}T_e, \quad (5)$$

$$r_s = \frac{1}{2}\frac{g_e R_e^2}{c_s^2}. \quad (6)$$

For application to coronal-hole jets in cool coronal holes, we chose $T_\odot = 1\times10^6$ K, for which $c_s = 1.3\times10^7$ cm s$^{-1}$ and $r_s = 5.8 R_\odot = 4.1\times10^{11}$ cm. We also set the mass density at radius $R_\odot$ to a low coronal-hole value, $\rho_\odot = 2\times10^{-16}$ g cm$^{-3}$, corresponding to a base pressure $P_\odot = 3.3\times10^{-2}$ dyn cm$^{-2}$. The density elsewhere is set by the condition of uniform mass flux, $\rho v_r r^2 = $ constant.

To simulate a jet originating in a compact minority-polarity region embedded in a quasi-uniform majority-polarity coronal hole, we superposed two initially potential magnetic fields. The locally uniform background field is represented by the Sun-centered monopole

$$\mathbf{B}_m = B_m \frac{R_e^2}{r^2}\hat{r}, \quad (7)$$

whose surface value $B_m = -2.5$ G. We positioned the small-scale embedded polarity at the equator (colatitude $\theta = \pi/2$) and central meridian ($\phi = 0$) of our spherical coordinate system, to optimize its resolution on the grid. It is represented by a point dipole, oriented in the radial direction and centered at depth $d$ below the surface,

$$\mathbf{B}_d = B_d d^3 \left\{\frac{\sin\theta\cos\phi}{D^3} - \frac{3}{2}\frac{r(R_e - d)[\cos^2\theta\cos^2\phi + \sin^2\phi]}{D^5}\right\}\hat{r}$$
$$+ B_d d^3 \cos\theta\cos\phi\left\{\frac{1}{D^3} - \frac{3}{2}\frac{r[r - (R_e - d)\sin\theta\cos\phi]}{D^5}\right\}\hat{\theta} \quad (8)$$
$$- B_d d^3 \sin\phi\left\{\frac{1}{D^3} - \frac{3}{2}\frac{r[r - (R_e - d)\sin\theta\cos\phi]}{D^5}\right\}\hat{\phi},$$

where

$$D^2 \equiv r^2 - 2r(R_e - d)\sin\theta\cos\phi + (R_e - d)^2 \quad (9)$$

and the peak radial field at the surface is $B_d = +35$ G. These superposed fields have the same ratio $B_d/B_m = -14$ as those assumed in our Cartesian jet studies. We set the depth $d = 1\times10^9$ cm, which positions the null point at height $h = 1.5\times10^9$ cm above the surface and the circular footprint of the dome-shaped separatrix surface at radius $a = 2.2\times10^9$ cm from the center of the embedded polarity.

The geometry of the configuration is shown in Figure 1. Selected magnetic field lines are drawn from one array of points just inside the separatrix surface (black curves), representing the outermost lines that form closed loops, and from a second array of points just outside that surface (white curves), representing the innermost lines that open to the heliosphere at their remote ends. Field lines rooted farther inside the circular footprint of the separatrix surface form shorter closed loops inward to the polarity inversion line (PIL) where $B_r$, which is color-shaded on the base plane of Figure 1a, changes sign (at the magenta/dark-blue transition). Field lines rooted farther outside the separatrix footprint all open to the heliosphere at their remote ends, and at large distances from the embedded polarity approach the purely radial orientation of the monopole field.

Our plasma and magnetic-field parameters in the configuration fix the ratio of thermal and magnetic pressures

$$\beta \equiv \frac{8\pi P}{B^2}, \quad (10)$$

The surface values range from $7.8 \times 10^{-4}$ at the center of the embedded polarity to $1.3 \times 10^{-1}$ in the uniform background field. In general, our coronal plasma is low-beta everywhere except in the vicinity of the null point, where $\beta$ is large and the thermal pressure exerted by the plasma plays a critical role in the reconnection dynamics. This region is marked in Figure 1 by a $\beta = 25$ isosurface (red spheroid suspended between the closed and open field lines). Figure 2 displays the variation of $\beta$ (solid curve) along a radial ray centered on the embedded polarity ($\theta = \pi/2$, $\phi = 0$) and passing through the initial coronal null point, along with the Alfvén speed $v_A$ (dashed curve) along the ray.

To energize the magnetic field that powers the jet, we imposed a slow rotational motion of the surface plasma within the PIL of the embedded dipole. The incompressible flow follows the contours of the radial magnetic field $B_r$ and so does not disturb the surface flux distribution. This choice maintains the initial magnetic field, prescribed above, as the reference potential field throughout the simulation, thereby simplifying the calculation of the magnetic free energy and helicity. For the tangential flow $\mathbf{v}_b$ at the base, we adopted the spatial and temporal profiles from our Cartesian studies

$$\mathbf{v}_b = v_0 f(t) \lambda_b \frac{B_2 - B_1}{B_r} \tanh\left(\lambda_b \frac{B_r - B_1}{B_2 - B_1}\right) \hat{r} \times \nabla B_r \quad (11)$$

and

$$f(t) = \frac{1}{2}\left[1 - \cos\left(\pi \frac{t}{t_b}\right)\right], \quad t \leq t_b;$$
$$= 1, \quad t > t_b. \quad (12)$$

The flow is smoothly accelerated from rest to peak speed over the duration $t_b = 1000$ s. In Equation (11), we set $\lambda_b = 5$, $B_1 = 2.5$ G, $B_2 = 30$ G, and $v_0 = 5.5 \times 10^{12}$ cm$^2$ s$^{-1}$ G$^{-1}$. They yield a peak flow speed in our configuration of $v_b = 2.5 \times 10^6$ cm s$^{-1}$, about 20% of the coronal sound speed and only 0.4% of the peak Alfvén speed at the surface. As will be shown below, the imposed motion also is far slower than the resultant reconnection-driven jet outflows. Outside of the region where $B_r \in [B_1, B_2]$, we set $\mathbf{v}_b = 0$, so that the magnetic field is line-tied across the entire coronal base of our simulation. The total velocity magnitude $|\mathbf{v}|$ at the surface (including the solar-wind outflow $v_r$) is color-shaded in Figure 1b. It is small outside the circle where $B_r = B_1 = 2.5$ G and in the interior region around the center of the embedded polarity where $\nabla B_r$ becomes small, and it reaches a broad peak in the annular region where $B_r \approx 10$ G. The imposed flow $\mathbf{v}_b$ is in the clockwise direction and imparts a right-handed twist to the field.

Two key global diagnostics are the volume-integrated magnetic and kinetic energies,

$$M = \iiint_V \frac{1}{8\pi} B^2 dV, \quad (13)$$

$$K = \iiint_V \frac{1}{2} \rho v^2 dV. \quad (14)$$

Particularly helpful are the magnetic free and excess kinetic energies, i.e. the increments attained beyond the steady-state values before the imposed footpoint flows are switched on:

$$\Delta M(t) = M(t) - M(0), \quad (15)$$

$$\Delta K(t) = K(t) - K(0). \quad (16)$$

To determine $M(0)$ and $K(0)$, we performed an initial relaxation simulation with no footpoint driving for $10^4$ s. Both $M$ and $K$ changed only minimally during this relaxation. We also calculated the time rate of change of total energy due to the Poynting flux,

$$\frac{dE}{dt} = \frac{1}{4\pi} \iint_S \left[ (\mathbf{v} \cdot \mathbf{B})\mathbf{B} - (\mathbf{B} \cdot \mathbf{B})\mathbf{v} \right] \cdot \hat{n} \, dS, \quad (17)$$

where the surface $S$ completely encloses the volume $V$ and the outward normal direction is $\hat{n}$. The principal contribution comes from the coronal base. Because the solar-wind inflow at this boundary is small and is aligned very closely with the local magnetic field ($\mathbf{v}_w \parallel \mathbf{B}$), it makes a negligible contribution to the Poynting-flux integral. The imposed tangential flow $\mathbf{v}_b$ injects nearly all of the additional energy, at the rate (note that $\hat{n} = -\hat{r}$)

$$\left. \frac{dE}{dt} \right|_{r=R_e} \approx -\frac{1}{4\pi} R_e^2 \iint (\mathbf{v}_b \cdot \mathbf{B}) B_r \sin\theta \, d\theta \, d\phi. \quad (18)$$

As will be shown below, prior to jet onset the magnetic free energy $\Delta M(t)$ closely tracks the time integral of the expression above.

A third useful global diagnostic is the relative magnetic helicity $H$. We employed the gauge-invariant expression (Finn & Antonsen 1985)

$$H = \iiint_V (\mathbf{A} + \mathbf{A}_p) \cdot (\mathbf{B} - \mathbf{B}_p) dV, \quad (19)$$

where $\mathbf{A}$ and $\mathbf{B}$ are the vector potential and magnetic field in the configuration at any time $t$, and $\mathbf{A}_p$ and $\mathbf{B}_p$ are the same quantities for the corresponding current-free field. As stated previously, the initial $\mathbf{A}_p$ and $\mathbf{B}_p$ provide the zero-helicity baseline at all times because the normal magnetic field at the coronal base does not change during the evolution. The time-invariant $\mathbf{A}_p$ and instantaneous $\mathbf{A}$, respectively, are given by

$$\mathbf{A}_p(r,\theta,\phi) = -B_m \frac{R_e^2}{r} \frac{\cos\theta}{\sin\theta} \hat{\phi} - \frac{1}{2} B_d d^3 \frac{r\sin\phi}{D^3} \hat{\theta} - \frac{1}{2} B_d d^3 \frac{r\cos\theta\cos\phi}{D^3} \hat{\phi} \quad (20)$$

and

$$r\mathbf{A}(r,\theta,\phi,t) = R_e \mathbf{A}_p(R_e,\theta,\phi) - \int_{R_e}^{r} dr' r' [\hat{r} \times \mathbf{B}(r',\theta,\phi,t)]. \quad (21)$$

We also calculate the time rate of change of helicity from the helicity flux,

$$\frac{dH}{dt} = \iint_S \left[ 2(\mathbf{A} \cdot \mathbf{v})\mathbf{B} - 2(\mathbf{A} \cdot \mathbf{B})\mathbf{v} - \frac{\partial \mathbf{A}}{\partial t} \times (\mathbf{A} - \mathbf{A}_p) \right] \cdot \hat{n}\, dS. \quad (22)$$

Its principal contribution comes from the coronal base, where $\mathbf{A} = \mathbf{A}_p$ and, as above, $\mathbf{v}_w \parallel \mathbf{B}$, so that only the $\mathbf{v}_b$ contribution to the first term is important. The result is

$$\left.\frac{dH}{dt}\right|_{r=R_e} \approx -2R_e^2 \iint (\mathbf{A}_p \cdot \mathbf{v}_b) B_r \sin\theta\, d\theta\, d\phi. \quad (23)$$

Our imposed footpoint motions inject positive net magnetic helicity into the corona. A numerical evaluation of the above integral on the simulation grid yields

$$\left.\frac{dH}{dt}\right|_{r=R_e} \approx 5.3 \times 10^{35} f(t)\, \text{Mx}^2\, \text{s}^{-1}. \quad (24)$$

As will be shown later, the time integral of this expression agrees very well with the numerically evaluated volume helicity $H$.

As already described, the tangential velocity $\mathbf{v}_b$ is prescribed everywhere on the coronal base, at $r = R_\odot$. The normal (radial) velocity $v_r$ is allowed to float freely, to accommodate wind outflow in open-field regions and quasi-static conditions in closed-field regions. In the guard cells below the open $r = R_\odot$ boundary, the mass density $\rho$ (therefore also the pressure) is held fixed at its Parker steady wind value. In the guard cells beyond the open $r = 9R_\odot$ boundary, we extrapolate the mass density using the ratio of outer to inner values in the Parker steady wind. All three velocity components are extrapolated there assuming zero-gradient conditions (free flow-through, free slip). The side boundaries in $\theta$ and $\phi$ are closed, reflecting with respect to the normal flow velocities ($v_\theta$ and $v_\phi$, respectively), and free slip with respect to the tangential flow velocities ($v_\phi$ and $v_\theta$, respectively, and $v_r$ in both cases). The mass density is extrapolated using zero-gradient conditions at all sides. The vector magnetic field $\mathbf{B}$ is extrapolated beyond all six boundaries assuming zero gradients.

The computational domain is a spherical wedge extending over $[1R_\odot, 9R_\odot]$ in radius $r$ and $[-9°,+9°]$ in both colatitude $\theta$ (centered at the equator) and longitude $\phi$ (about the central meridian). This amounts to a 25:1 aspect ratio in total distance spanned along and across the radial direction, and the radial extent is sufficient to place the outer boundary well beyond the sonic point $r_s$. The grid was spaced uniformly in angle and exponentially in radius, so that $\Delta r/r$ is uniform everywhere. A 35×5×5 grid of root blocks (with 8×8×8 cells per block) coarsely resolved the domain into 280×40×40 approximately cubic cells. Up to four additional levels of grid refinement, by a factor of two at each level, were allowed during the simulation. The entire coronal base, extending to $5\times10^8$ cm in height, was required to refine to the maximum level, corresponding to an area resolution of 640×640. The same maximum refinement was imposed on a central region encapsulating the null point and separatrix dome, reaching $2.2\times10^9$ cm in height and extending over $[-2.5°,+2.5°]$ in angle. The outermost perimeter of root blocks (within 3.6° of the side boundaries) was restricted to only one additional level of refinement beyond a height of $2.2 \times 10^9$ cm above the surface, to avoid overly resolving regions that are near the side boundaries and far from the central, jet-containing portion of the domain. The resulting initial grid consisted of roughly $2 \times 10^4$ blocks ($10^7$ cells), about 0.5% of that needed to fully resolve the entire domain. The maximum resolution is $5 \times 10^{-4} R_\odot = 3.5 \times 10^7$ cm at and near the surface. A close-in view of the initial grid is shown in Figure 3a.

In order to refine the evolving current structures at the null point, separatrix surface, and the eventual jet, we tested the local strength of the current density and magnetic field at every grid point at 25 s intervals as in our previous investigation of solar eruptions (Karpen et al. 2012). Specifically, we evaluated

$$c \equiv \frac{\left|\iint_S \nabla \times \mathbf{B} \cdot d\mathbf{a}\right|}{\oint_C |\mathbf{B} \cdot d\mathbf{l}|} = \frac{\left|\oint_C \mathbf{B} \cdot d\mathbf{l}\right|}{\oint_C |\mathbf{B} \cdot d\mathbf{l}|} = \frac{\left|\sum_{l=1}^{4} B_l L_l\right|}{\sum_{l=1}^{4} |B_l L_l|} \quad (25)$$

and

$$b \equiv \frac{\oint_C |\mathbf{B} \cdot d\mathbf{l}|}{\oint_C dl} = \frac{\sum_{l=1}^{4} |B_l L_l|}{\sum_{l=1}^{4} L_l}. \quad (26)$$

$S$ is a grid face through which any component of current passes, $C$ is a contour bounding $S$, $L$ is a discrete segment of $C$ on the grid ($l = 4$ segments per contour), and $B_l$ is the tangential component of $\mathbf{B}$ along $L$. The dimensionless current measure $c$ ranges from 0 for a potential field to 1 for a field carrying maximal current. We marked a block to be refined when $c$ exceeded 0.1 in any cell within the block, and to be coarsened when $c$ fell below 0.025 in every cell within it. We also required that the average tangential field strength $b$ be no greater than 0.25 G in the cells marked for refinement. This combined test targeted null points and relatively highly twisted magnetic field lines, in particular those supporting the nonlinear Alfvén waves. In response to this test, the simulation grid expanded to well over $2.5 \times 10^5$ blocks, or $1.3 \times 10^8$ cells, after the jet was initiated and became increasingly extended into the outer corona of our domain. A mid-range view of the final grid is shown in Figure 3b.

## III. Results

The evolution of our configuration features three principal phases: (1) a long interval of gradual energy buildup; (2) a short, explosive episode of instability-driven reconnection onset, energy release, and jet generation; and (3) an extended propagation phase that begins with the jet generation but continues after the energy release has concluded and the jet front travels away toward the heliosphere. This sequence mirrors the evolution of our first Cartesian simulation of a coronal jet (P09). For clarity, we note that the third phase was characterized differently in that paper: it begins when the energy release concludes, and it exhibits a relaxation of the residual magnetic stresses toward a new, quasi-force-free state in the inner corona. We also observe that evolution in our simulation, but our focus here is the ongoing jet propagation into the outer corona and heliosphere.

The first two phases of evolution of the jet source region in the inner corona are illustrated by Figures 4–6 and the accompanying Movies 1 and 2. The third phase, when the jet transits the corona, is illustrated by Figures 7 and 8 and the accompanying Movies 3 and 4. The time intervals covered by the three episodes are approximately $t \in [0\,\text{s}, 2750\,\text{s}]$, $t \in [2750\,\text{s}, 3500\,\text{s}]$, and $t \in [2750\,\text{s}, 4000\,\text{s}]$, respectively. We point out that the jet is continuing to propagate outward when the simulation is terminated at $t = 4000$ s to avoid outer-boundary influences on the jet front.

During the energy buildup, shown in Figure 4 and Movie 1, the imposed clockwise tangential flow $\mathbf{v}_b$ within the embedded polarity introduces a counter-clockwise twist to the closed lines of magnetic field $\mathbf{B}$ in that region. The induced twist component of magnetic field increases the total field strength and pressure within the separatrix dome. This causes the dome to expand, primarily upward, to maintain force balance with its surroundings. The axisymmetry of the initial configuration is maintained robustly until very late in this phase, while the initial low hemispherical dome (Fig. 4a) extends to become a prolate tower (Fig. 4d). As the isosurface of $\beta = 2$ (green spheroid) in the figure indicates, the null point at the top of the dome nearly triples its height as the free energy accumulates below. The maximum number of turns of twist, $N$, induced by the motion occurs at radius $a_b \approx a/4 = 5.5 \times 10^8$ cm, where the tangential flow attains its peak speed $v_b = 2.5 \times 10^6$ cm s$^{-1}$. After $\Delta t_b = 2000$ s of displacement at this speed (note that the average speed is $v_b/2$ during the 1000 s ramp-up interval), the accumulated maximum twist at $t = 2500$ is

$$N = \frac{v_b \Delta t_b}{2\pi a_b} \approx \frac{5}{1.1\pi} \approx 1.45 \ . \quad (27)$$

Throughout this phase, the kinetic energies due to the imposed tangential flow and the upward expansion of the separatrix dome remain negligible, whereas the magnetic energy increases substantially, as is shown in Figure 9.

As the energy buildup phase concludes, the evolution transitions suddenly to an explosively dynamic phase of instability onset, initiation of fast reconnection, rapid conversion of magnetic energy into kinetic energy, and generation of the coronal jet. Close examination of the plasma-$\beta$ isosurface in Figure 4d shows that the symmetry of the separatrix dome is beginning to break at $t = 2525$ s. As can be seen best in Movie 1, the separatrix tower buckles to the left and then begins to rotate counter-clockwise in an unwinding motion. The null-point volume promptly flattens, spreads into a current patch, and fragments in response to the buckling of the tower, as shown in Figure 5a. It then rotates around the axis of the jet source region along with the convulsing magnetic field, from left to front to right to rear in the figure, and thereafter at a progressively slower rate as the rotation slows and stops. Overall, the tower and its magnetic field lines unwind roughly the 1.5 turns of twist that were introduced by the footpoint motion during the energy buildup.

This transition from quasi-static buildup to explosive release has been identified as resulting from an ideal kink-like instability of the twisted magnetic field inside the separatrix tower (P09). The buckling of the tower at a critical number of turns of twist was replicated under magnetically force-free evolution by a perfectly ideal, Lagrangian simulation model, FLUX (Rachmeler et al. 2010). Unlike the finitely resistive ARMS simulations, however, the ideal FLUX calculation did not unwind the buckled tower and no energy was released. These key aspects of the ARMS MHD evolution depend critically upon fast reconnection driven by the instability-induced convulsion of the magnetic field, as discussed at length in P09.

The unwinding of the magnetic field occurs as twist stored on the closed field lines of the separatrix tower is transferred to untwisted open field lines of the coronal hole by reconnection across the current patch formed by the distorted null. This can be seen best in Movie 1, although it also is suggested by the sequence of still images in Figure 5. Whereas the open field lines exiting the top of the system are all laminar and monotonically spaced in Figure 4, in Figure 5 they are strongly bent and tangled above and adjacent to the separatrix tower. This change reflects the transient twist imparted to those open lines by their reconnection with closed lines within the tower. The transient twist then propagates away toward the outer corona and inner heliosphere, leaving the lower portions of these open field lines to relax back toward a more quiescent and laminar state resembling that shown in Figure 4.

The onset of instability-driven reconnection radically restructures the electric current density within the jet source region. Isosurfaces of current-density magnitude ($|\mathbf{J}|/c = 1.4\times10^{-9}$ G cm$^{-1}$) are shown in Figure 6, transitioning from the end of the energy-buildup phase through most of the energy-release phase. The accompanying Movie 2 shows the entire evolutionary sequence. Before the separatrix tower buckles (Fig. 6a), the isosurface approximates a vertical cylinder confined to the interior. This volumetric current is associated with the broadly distributed magnetic twist within the tower. After the tower buckles (Figs. 6b-d), strong currents rapidly form at the separatrix and cover a large fraction of this surface. This surface current is due to the strong misalignments between the twisted magnetic field inside the tower and the quasi-potential untwisted field outside, in locations where they are driven together by the ideal convulsion. The redistributed current fragments as localized patches strengthen and then weaken during the precession of the reconnection region around the circumference of the tower.

The twist imparted to the reconnected open field lines generates a nonlinear Alfvén wave that powers plasma outflow with its wave pressure. The compressed and accelerated plasma trails behind the leading wave front, flowing outward at speeds well below the Alfvén speed (P16). These features are illustrated by mid-range views of total velocity magnitude and logarithm of mass density during the jet propagation phase in Figures 7 and 8, respectively. The accompanying Movies 3 and 4 show the full evolutionary sequences. At $t = 2750$ s (Fig. 7a) the velocity magnitude is dominated by the background solar wind, except at the separatrix tower where the jet generation is just beginning. Over the entire domain, the wind speed $v_w < 2\times10^7$ cm s$^{-1}$, whereas the color scale in the figure saturates at $|\mathbf{v}| = 5\times10^7$ cm s$^{-1}$. The high speed of the jet outflow strongly hints at its Alfvénic character, which we demonstrate more quantitatively below. Because the reconnection region precesses around the axis of the tower as the twist unwinds, the jet as a whole acquires a helical structure. This is indicated by the location of regions with the highest speed (magenta shading) at the upper side of the fan-shaped jet volume in Figure 7b and at its lower side in Figure 7c. We also observe clear evidence for a fast magnetosonic wave propagating across the corona as the jet broadens and then narrows in Figure 7b-d. This wave causes the bent portions of the field lines to disappear from (Fig. 7c), then reappear in (Fig. 7d), the field of view at intermediate radii, as is easily seen in Movies 3 and 4.

Figure 8 and Movie 4 show the evolution of the mass density during the jet propagation. As late as $t = 3150$ s (Fig. 8b), the mass density is dominated by the stratification associated with the solar wind. Subsequently (Figs. 8c-d), the outflow of compressed plasma becomes quite evident, all of the way to the outer edge of this field of view, $r = 2.5\ R_\odot$. Due to the spatially and temporally intermittent nature of the reconnection that drives the Alfvén wave and jet outflow, the density structures are highly filamentary, as can be seen in the figure and the movie. The last frame (Fig. 8d) shows that the outflow continues in the inner corona, in response to the sustained reconnection and jet generation, long after the leading front has passed into the outer corona, bound for the heliosphere. Consequently, the jet develops a very extended structure along its direction of propagation.

In Figure 9, we display the time histories of magnetic free energy $\Delta M$ and excess kinetic energy $\Delta K$, defined in Equations (15)–(16), as solid and dashed curves, respectively. The first two episodes of energy buildup and energy release are readily evident in this figure. During the energy buildup, from $t = 0$ to ~2750 s, the imposed tangential flow $\mathbf{v}_b$ ramps up from rest at $t = 0$ to attain its peak speed at $t_b = 1000$ s, as prescribed by Equation (12). Throughout this phase, the excess kinetic energy – the amount above the ambient kinetic energy in the solar wind – remains negligibly small, as seen in the figure. Meanwhile, the magnetic free energy increases slowly at first, during the ramp-up to $t = t_b$ (marked by the vertical dotted line), and rises essentially linearly with time thereafter, when the flow is maintained at fixed speed. This change in magnetic energy tracks very closely the time integral of the Poynting flux through the base of the domain, defined in Equation (18), as the tangential flow slowly twists and energizes the field below the separatrix dome.

The second phase (energy release) begins with the simultaneous attainment of a maximum in the magnetic free energy $\Delta M$ and onset of a steep increase in the excess kinetic energy $\Delta K$, at $t \approx 2750$ s. During the remainder of this phase, through $t \approx 3500$ s, the magnetic energy decreases and the kinetic energy increases steeply, signaling a rapid conversion of magnetic to kinetic energy. As described above, this magnetic-energy release is accompanied by the onset of impulsive reconnection and the generation of the jet. As this phase ends, the magnetic free energy reaches a minimum while the excess kinetic energy attains a maximum. The peak values during this phase are $\Delta M \approx 9.0 \times 10^{28}$ erg and $\Delta K \approx 1.7 \times 10^{28}$ erg at the beginning and end, respectively. Thereafter, $\Delta M$ resumes its upward trajectory due to the ongoing footpoint motions, which add new twist to the closed field lines. Meanwhile, $\Delta K$ temporarily decreases, then initiates a new trend of rising at a much slower rate than that in the energy-release phase. Its slope is close to that of the magnetic free energy at late times $t > 3750$ s, just prior to the end of the simulation. In addition to winding up the magnetic field, the footpoint motions keep driving slow outflows behind the jet, through continued reconnection between the twisting field of the embedded bipole and the external open flux. The key point is that, prior to the jet, the system maintained its axisymmetry to a very high approximation and, consequently, no reconnection was possible. After the jet, however, the symmetry is forever lost and any subsequent energy buildup drives continued slow reconnection at the null and separatrix prior to the onset of any subsequent jet (e.g., P10). In a more recent simulation (Roberts et al. 2016), we ramped down the footpoint motions as the reconnection began. In that case, both the magnetic free and excess kinetic energies leveled off at the conclusion of the energy-release phase.

To quantify the transfer of energy from the inner to the outer corona, we evaluated the full Poynting flux integral in Equation (17) over $\theta$ and $\phi$ at several different radii $r$. The time integral of those fluxes up to time $t$ yields the magnetic energy injected into the coronal volume above those heights. We display the results at $r/R_\odot = 1.5$, 2.0, 3.0, and 4.0 in Figure 10. The curves at 1.5 and 2.0 seem to be saturating by the end of the simulation, at $t = 4000$ s, at values above $2 \times 10^{28}$ erg. The curves at 3.0 and 4.0 $R_\odot$ are still rising, but we cannot state with assurance what would transpire further from the Sun. It seems plausible that nearly dissipation-free transfer of magnetic energy will persist well into the outer corona, consistent with our displayed results for $r/R_\odot$ = 1.5 and 2.0. A more extended simulation on a larger domain, with footpoint motions ramped down by the time of jet onset (Roberts et al. 2016), is required for us to reach a definitive conclusion, however.

For completeness, we also calculated but do not show the time integrals of the excess fluxes of enthalpy and kinetic energy (above and beyond those carried by the ambient solar wind). The integrals exhibit similar time delays and relative amplitudes as those of the Poynting flux integrals shown in Figure 10, but overall they are an order of magnitude smaller. Thus, the total flux of energy into the corona in our simulation is dominated by the Poynting flux of magnetic energy shown in the figure.

The relative magnetic helicity given in Equation (19) measures the amount and disposition of magnetic twist in our configuration. We evaluated the direct volume integral and also the time integral of the helicity flux through the coronal base, given in Equation (24). As shown in Figure 11, both values increase monotonically throughout the simulation. No helicity is lost from the domain, as the propagating disturbances did not reach the outer boundary before the simulation ended. Indeed, we find that the helicity is better than conserved, in that the volume integral exceeds the time-integrated surface integral late in the simulation, though only by a small percentage. This error of excess evidently is due to the exponential stretching of the spherical grid, which causes the grid spacing to increase with radius and reduces the accuracy of the numerical volume integration. In a companion Cartesian simulation of this system without gravity and wind (DeVore et al. 2016), the agreement between the volume and surface integrals is much better: +1.5% error in the Cartesian vs. +7.5% in the spherical setup. We found exactly the same trend in errors between the volume magnetic free energies and the integrated Poynting fluxes at the coronal base in the two simulations.

Despite the small quantitative error in the helicity, its qualitative behavior is striking and informative. There is no signature of any significant change in the helicity evolution in Figure 11 at the time of reconnection onset and initiation of the jet, $t \approx 2750$ s. Therefore, very little diffusive slippage of magnetic field through the plasma occurs during the extensive rearrangement of connectivity that follows. Consequently, the evolution in our simulation is predominantly, if not entirely, due to breaking and reconnecting of field lines, rather than diffusion. This result implies that our simulation is accurately capturing the evolution of the highly electrically conducting corona.

Using the profiles of Alfvén speed and solar-wind speed along the radial ray at $\theta = \pi/2$, $\phi = 0$, we have performed a simple quantitative analysis of the propagation of our wave front. The propagation time $\tau_p$ for an Alfvén wave launched at radius $r_m$ to travel to radius $r$ is

$$\tau_p = \int_{r_m}^{r} \frac{dr'}{v_A(r') + v_w(r')}. \quad (28)$$

As the initiation point, we chose the radius $r_m \approx 1.09$ where the Alfvén speed attained its minimum value at the jet onset time $t = 2750$ s, as shown in Figure 12a. The result for the propagation time $\tau_p$ is shown as the solid curve in the Figure 12b. For comparison, we inspected the same radial ray at selected later times and identified the outermost extremum in the transverse magnetic field strength $|\mathbf{B}_\perp|$. The radial positions of those extrema are plotted as filled circles in the figure. Clearly, the wave front leading the jet is propagating at a speed very close to the local Alfvén speed in a frame moving with the solar wind. This predicted trajectory is followed for the entire duration (1250 s) of the propagation phase of our jet as its front traverses the corona ($4R_\odot$) on its way to the heliosphere.

## IV. Analysis

In this section, we summarize some key features of our simulated jet and discuss their application to coronal-hole jets observed on the Sun. First, we consider the nature of the energy buildup, the rise in height of the closed separatrix dome, and the ideal instability that eventually convulses the dome. Then, we note that each phase of the evolution exhibits its own distinct, characteristic time scale: for energy buildup, jet generation, and jet propagation. These time scales, plus the jet length and speed of transverse motion, are determined by coronal parameters such as the size of the source region, the rate of footpoint driving, the local Alfvén speed, and the Alfvén speed in the outer corona as the jet propagates toward the heliosphere.

An important issue for our jet model is the nature of the ideal instability that triggers the burst of reconnection. Figures 4 and 9, which show how the energy builds up, yield considerable insight into this issue. We note from Figure 4 that the main effect of the photospheric stressing is to increase the height, $h$, of the closed-field region, rather than its radius, $a$. This is to be expected, because the vertical lines of the open background field strongly resist the horizontal expansion of the closed flux, especially near the base where the line-tied photosphere prevents any expansion whatsoever. In the vertical direction, in contrast, the field strength falls off to vanish at the null, so the closed flux is relatively free to expand upward. Figure 9 indicates that this expansion results in a linear increase in the magnetic free energy, $\Delta M$, with time after the initial ramp-up. The free energy is due primarily to the twist field, $B_\phi$, in the closed region,

$$\Delta M = \pi a^2 h B_\phi^2 \propto t. \quad (29)$$

On the other hand, the twist flux clearly increases linearly with time for fixed-speed photospheric rotation,

$$\Phi_\phi = ah B_\phi \propto t, \quad (30)$$

which accounts for the linear buildup of helicity seen in Figure 11. Because $a$ is fixed, the only solution to these two constraint equations is for $B_\phi$ to remain constant while $h$ increases linearly with time during the evolution. Indeed, careful examination of the position of the null during the energy-buildup phase in Figure 4 shows that it rises upward at a steady speed.

These results also shed light on the physical nature of the ideal instability. The initial closed flux forms a hemispherical structure with radius $a$ roughly equal to height $h$. As the energy builds up, however, the closed flux region becomes more and more columnar with increasing aspect ratio, $h/a$. Eventually, the aspect ratio becomes so large that the structure buckles, triggering the burst of reconnection. This intuitive picture also accounts for the finding in our previous work that the instability sets in earlier for greater inclination of the background open field (e.g., P10, P15). For an inclined field, the closed-flux tower already is tilted, and is more likely to buckle earlier.

We now turn to the time scales characterizing the three phases of evolution of our system, in order of occurrence. The duration of the energy-buildup phase, $\tau_b$, is determined by the speed of the imposed energizing flows and the size of the jet source region. For computational convenience, in our simulation we imposed flows that are artificially fast relative to observed random surface motions on the Sun. The instability was initiated after $N \approx 1.45$ turns of twist were accumulated along a circular path of radius $a_b \approx a/4$, where $a \approx 2\times10^9$ cm is the radius of the separatrix dome, at a (peak) speed $v_b \approx 2.5\times10^6$ cm s$^{-1}$. The resulting characteristic time $\tau_b$ is

$$\tau_b = \frac{2\pi N a_b}{v_b} \approx \frac{2\pi N(a/4)}{v_b} = \frac{\pi N}{2}\frac{a}{v_b} \approx 2.5\frac{a}{v_b}. \quad (31)$$

A numerical evaluation gives $\tau_b \approx 2\times10^3$ s, or about 0.5 hr, as we found in our simulation. This happens to be on the order of the recurrence periods of some observed homologous jets, whose quasi-periodicity has been ascribed to ongoing flux emergence and cancellation in the source regions (e.g., Cheung et al. 2015 and references therein). Our model does not represent these processes, so the agreement is wholly fortuitous; furthermore, the observed speeds of photospheric flux-emergence and -cancellation flows are well below $2.5\times10^6$ cm s$^{-1}$. If we were to repeat our experiment at typical random photospheric flow speeds ($v_b \approx 10^5$ cm s$^{-1}$), the energy-buildup time $\tau_b$ would be 25 times longer, i.e. about 12 hr. If the photospheric flows were less organized, or distributed more broadly across the source ($a_b > a/4$), then $\tau_b$ would be even longer, on the order of 1 day or more. These estimates suggest that jets should occur randomly several times per day in coronal holes, in proportion to the number of source regions that are present, but they should not occur routinely in every source multiple times per day. This prediction of our model seems to be in broad general agreement with the rates of jet occurrence that are observed.

In contrast to the energy-buildup phase, which is driven by imposed boundary motions, the characteristic time scales for the jet-generation and jet-propagation phases are intrinsic to the coronal dynamics. They are wholly independent of the driving speed $v_b$ and the preceding time scale $\tau_b$ for energy buildup. As has been shown here and in P09, the jet-generation scale $\tau_g$ is the time required for the region of newly reconnecting field lines to sweep completely around the separatrix dome, transferring interior magnetic twist onto open field lines where it can propagate into the outer corona and heliosphere. This duration coincides approximately with the elapsed time between the onset of increase of kinetic energy in the jet and the attainment of its peak value. We found that the field untwists the same number of turns ($N \approx 1.45$) as were acquired in the energy-buildup phase, and that the reconnection region traverses an intermediate radius $a_g \approx a/2$ at the side of the dome after the twisted column buckles. The speed $v_g$ of precession of the reconnection region about the dome is determined by the rate of inflow into the current sheet. Numerical studies of magnetic reconnection suggest that this inflow speed is roughly 15% of the Alfvén speed at the side of the sheet, i.e. at the surface of the dome, $v_{AD}$. Putting all of this together, we deduce that the characteristic jet-generation time is

$$\tau_g = \frac{2\pi N a_g}{v_g} \approx \frac{2\pi N(a/2)}{0.15 v_{AD}} \approx 20 N \frac{a}{v_{AD}} \approx 30 \frac{a}{v_{AD}}. \quad (32)$$

By inspection, we find that locally $v_{AD}$ is about $7\times10^7$ cm s$^{-1}$ at the surface of the dome; it is somewhat higher inside where the twist is greater and the mass density is lower. Numerically evaluating the result above, we find $\tau_g \approx 9\times10^2$ s, or about 15 min. This is very close to the duration of the kinetic-energy rise (750 s), and fully consistent with the directly measured precession of the reconnection region in our simulation. It also agrees well with the typical duration of observed coronal jets (e.g., Shimojo et al. 1996; Cirtain et al. 2007; Savcheva et al.

2007). The direct dependence of $\tau_g$ on the dome size, $a$, and its inverse dependence on the Alfvén speed, $v_{AD}$, allow for a range of observed jet durations on the Sun.

We point out that the relationship given above can be combined with measurements of the duration $\tau_g$ and dome size $a$ in an observed jet to derive an independent estimate for the local Alfvén speed $v_{AD}$. Because the wave front is Alfvénic, the leading edge of the jet progresses at the speed $v_{AD}$, and this speed also can be measured directly. Our value, $v_{AD} \approx 7\times10^7$ cm s$^{-1}$, agrees very well with observations of *Hinode* X-ray jets (Cirtain et al. 2007) and of *SOHO* and *STEREO* white-light jets (e.g., Wang et al. 1998; Wood et al. 1999; Yu et al. 2014), which indicate that the speed of the leading front lies between $4\times10^7$ cm s$^{-1}$ and $1\times10^8$ cm s$^{-1}$ in the low corona. In addition, our model predicts the length of the jet given the speed of its front $v_{AD}$ and its duration $\tau_g$,

$$L_{jet} = v_{AD}\tau_g \approx 30a. \quad (33)$$

In our case, we find $L_{jet} \approx 6\times10^{10}$ cm, well within the range of lengths of observed coronal-hole jets (e.g., Shimojo et al. 1996; Cirtain et al. 2007; Savcheva et al. 2007).

The precession of the reconnection region around the separatrix dome also produces an apparent motion of the jet source across the plane of the sky. The projected linear distance traveled is smaller than the circumference by a factor $2/\pi$. Therefore, the apparent speed of motion across the sky, $w_g$, is

$$w_g = \frac{2}{\pi}v_g \approx \frac{2}{\pi}0.15v_{AD} \approx \frac{1}{10}v_{AD}. \quad (34)$$

For our jet, we find $w_g \approx 7\times10^6$ cm s$^{-1}$, which also agrees with the range of values observed on the Sun (Cirtain et al. 2007; Savcheva et al. 2007; Shimojo et al. 2007).

As the Alfvénic jet front propagates from the side of the dome through the outer corona on its way to the inner heliosphere, it attains the local Alfvén speed $v_{Ap}$ in the frame moving at the solar wind speed $v_w$. The wave front then traverses 1 $R_\odot$ during a time $\tau_p$ given by Equation (28), under the assumption of constant $v_{Ap}$ and $v_w$,

$$\tau_p = \frac{R_e}{v_{Ap}+v_w} \approx \frac{R_e}{v_{Ap}}. \quad (35)$$

For the average Alfvén speed $v_{Ap} \approx 2.5\times10^8$ cm s$^{-1}$ in our solar wind between about $1.5R_\odot$ and $9R_\odot$, we find that $\tau_p \approx 300$ s, or about 5 min per solar radius traveled. This agrees very well with the measured arrival times of the wave front in the outer corona of our simulation and confirms the Alfvénic nature of the jet front as it propagates toward the inner heliosphere. Under these conditions, the leading front of a well-positioned solar jet would reach *Solar Probe Plus* at perihelion ($r < 12R_\odot$) for in-situ sampling in less than 1 hr after onset in the low corona.

**V. Discussion**

In this paper, we describe and analyze a numerical simulation of a solar coronal-hole jet. Our study is based on the embedded-bipole model (Antiochos 1990, 1996) for jets in coronal holes, in which nearly discontinuous changes in magnetic-field direction are established readily between the closed field of the embedded bipole and the open field of the surrounding coronal hole. The onset of fast magnetic reconnection across an electric current sheet gives rise to impulsive solar activity in the corona, including the generation of coronal-hole jets. Previously, this model was verified through gravity- and solar-wind-free simulations in Cartesian geometry (P09, P10, P15, P16; Dalmasse et al. 2012). By including the effects of spherical geometry, solar gravity, solar wind, nonuniform ambient magnetic field, and density stratification of the atmosphere, we have performed a definitive test of our model under fully realistic conditions. Our results also demonstrate that the simulated jet propagates readily from the inner to the outer corona. Thus, we conclude that reconnection-driven jets from embedded bipoles in coronal holes persist into the inner heliosphere, bearing signatures that can be remotely sensed or sampled in situ. This result is consistent with well-established observations of white-light jets (§1). Our findings also provide support for the hypothesis that the small-scale microstreams observed in the solar wind are heliospheric manifestations of coronal-hole jets, and suggest a possible origin for the torsional Alfvén waves detected in the interplanetary medium.

We chose parameters for our simulation that are typical of coronal holes (ambient plasma density and temperature, and magnetic field strength) and their embedded bipoles (field strength and region size) for maximum realism. The properties of the resulting simulated jet (duration, velocity, length, energy) are all typical of those observed routinely on the Sun. Based on our results, we developed scaling relationships between jet properties and the inherent characteristics of the coronal source (region size, inner and outer coronal Alfvén speeds, driving speed of footpoint motions) that can be tested against observations and may explain the range of values exhibited by observed coronal-hole jets.

Additional simulation studies of our coronal-hole jet model are underway. First, we are investigating in greater detail the properties of the nonlinear Alfvén waves that are generated by reconnection onset in the inner corona and then propagate into the outer corona (DeVore et al. 2016). Second, we are developing an understanding of the magnetohydrodynamic turbulence aspects of these jets, with implications for both the theory of reconnection-driven turbulence and interplanetary observations (Uritsky et al. 2016). Third, it is important to understand the expected in-situ signatures of coronal-hole jets in order to prepare for the unprecedented science data anticipated from *Solar Probe Plus*. Thus, in a follow-up simulation, we are extending our simulation grid much farther into the heliosphere and are tracking the evolution of the jet over a much longer time span (Roberts et al. 2016).

We conclude by pointing out that the embedded-bipole model also yields reconnection-driven jets when the background magnetic field is a large-scale closed loop (Wyper & DeVore 2016), rather than an open coronal hole. Therefore, the basic reconnection-driven jet scenario suggested by Antiochos (1990, 1996) is robust and can account for both coronal-hole and active-region jets on the Sun. At sufficiently high numerical resolution in either scenario, the spatially and temporally intermittent character of the separatrix reconnection process produces numerous small-scale structures, including magnetic flux ropes, plasma threads and blobs, and current filaments (Wyper et al. 2016). These additional aspects of the generation of solar jets are being explored, and undoubtedly will have important implications for understanding the heating of the corona and the acceleration of the solar wind.

**Acknowledgements**

We thank Kevin Dalmasse and Peter Wyper for many helpful discussions of the results developed in this paper. Our research was supported by NASA Living With a Star grants to understand reconnection-driven solar jets and their heliospheric consequences, and to prepare for the investigation of the inner heliosphere by *Solar Probe Plus*. The numerical simulations were supported by grants to C.R.D. of High-End Computing resources at NASA's Center for Climate Simulation.

**Figures & Captions**

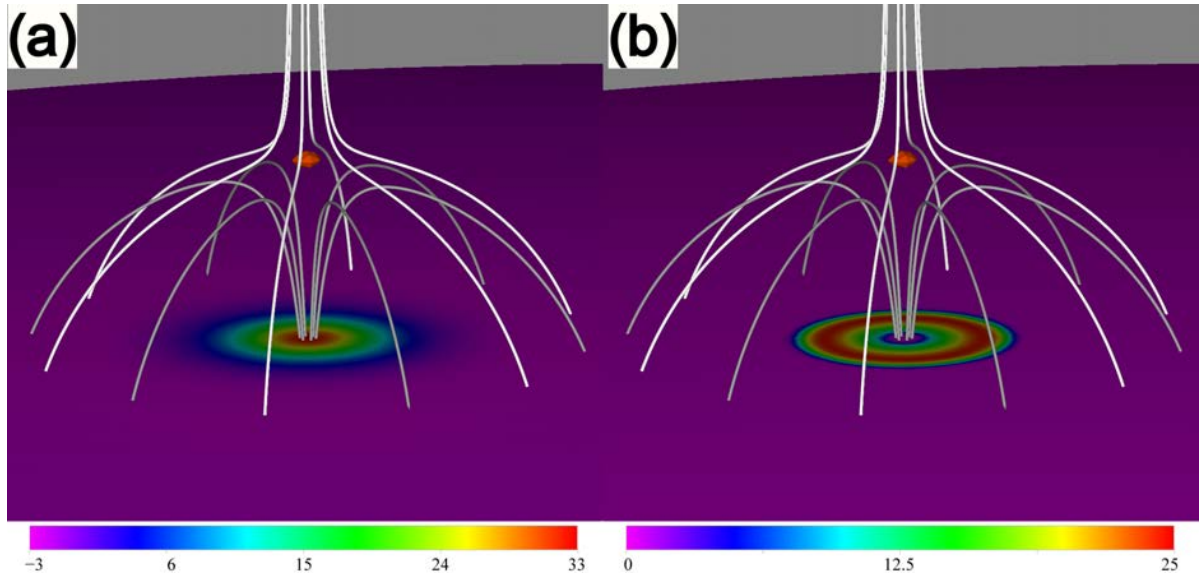

**Figure 1**. Perspective view of the jet source region. Magnetic field lines defining the separatrix surface either close to the Sun just inside the surface (gray curves) or open to the heliosphere just outside the surface (white curves). The magnetic null point is enclosed by an isosurface of plasma $\beta = 25$ (red spheroid). Color shadings across the coronal base indicate (a) the sign and magnitude of the radial magnetic field component $B_r$ (G) and (b) the magnitude of the plasma total flow velocity $\mathbf{v}$ ($10^5$ cm s$^{-1}$), whose horizontal component $\mathbf{v}_b$ is directed clockwise and is the dominant contributor to $\mathbf{v}$.

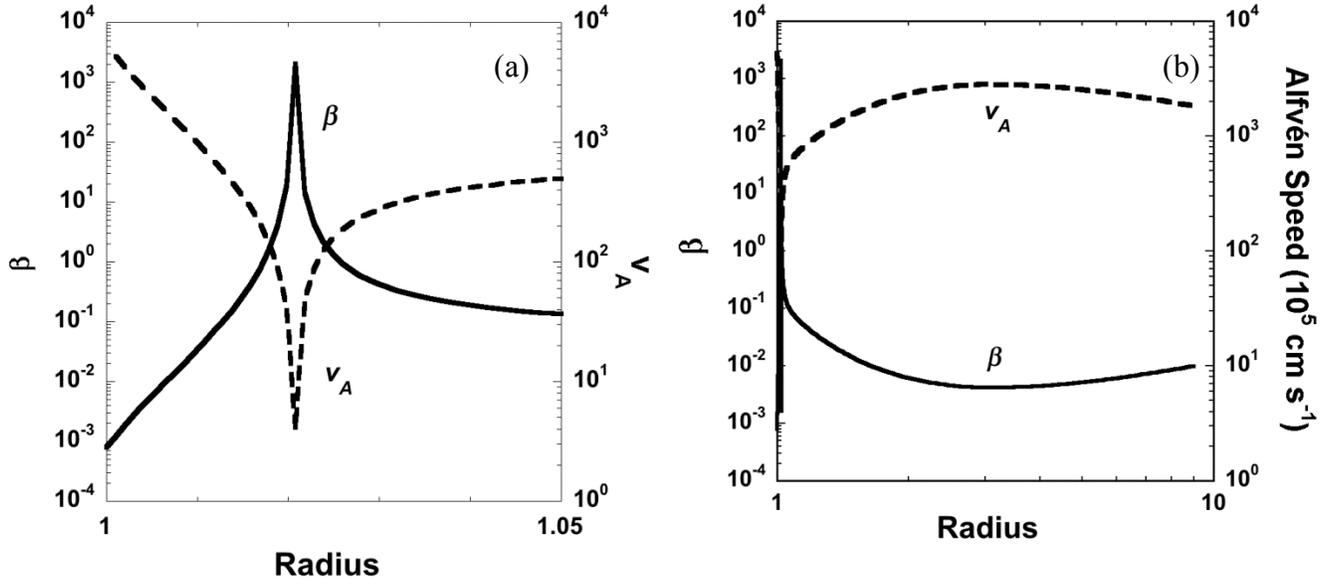

**Figure 2.** Initial plasma $\beta$ (solid curve) and Alfvén speed ($v_A$; dashed curve) vs. normalized radius $r/R_\odot$ along the radial ray at $\theta = \pi/2$, $\phi = 0$ that passes through the coronal null point: (a) close-in view. (b) full-range view.

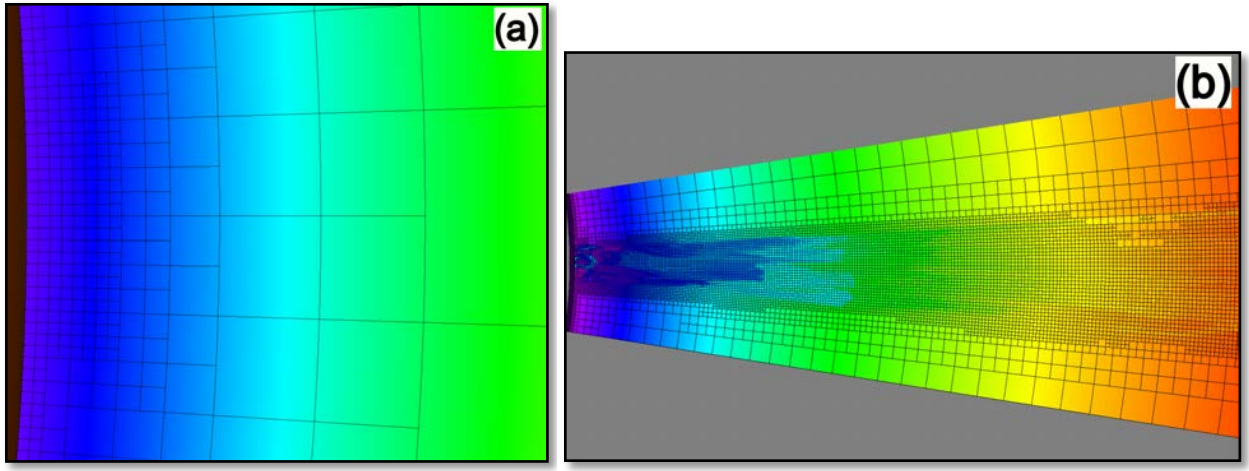

**Figure 3.** The adaptively refined grids in the equatorial ($\theta = \pi/2$) plane: a) close-in view at $t = 0$. (b) mid-range view at $t = 4000$ s. Black lines are boundaries of grid blocks; there are $8\times8\times8$ cells per block. The logarithm of the mass density $\rho$ (g cm$^{-3}$) is color-shaded in the plane of the sky. The coronal base is the dark surface to the left in each image.

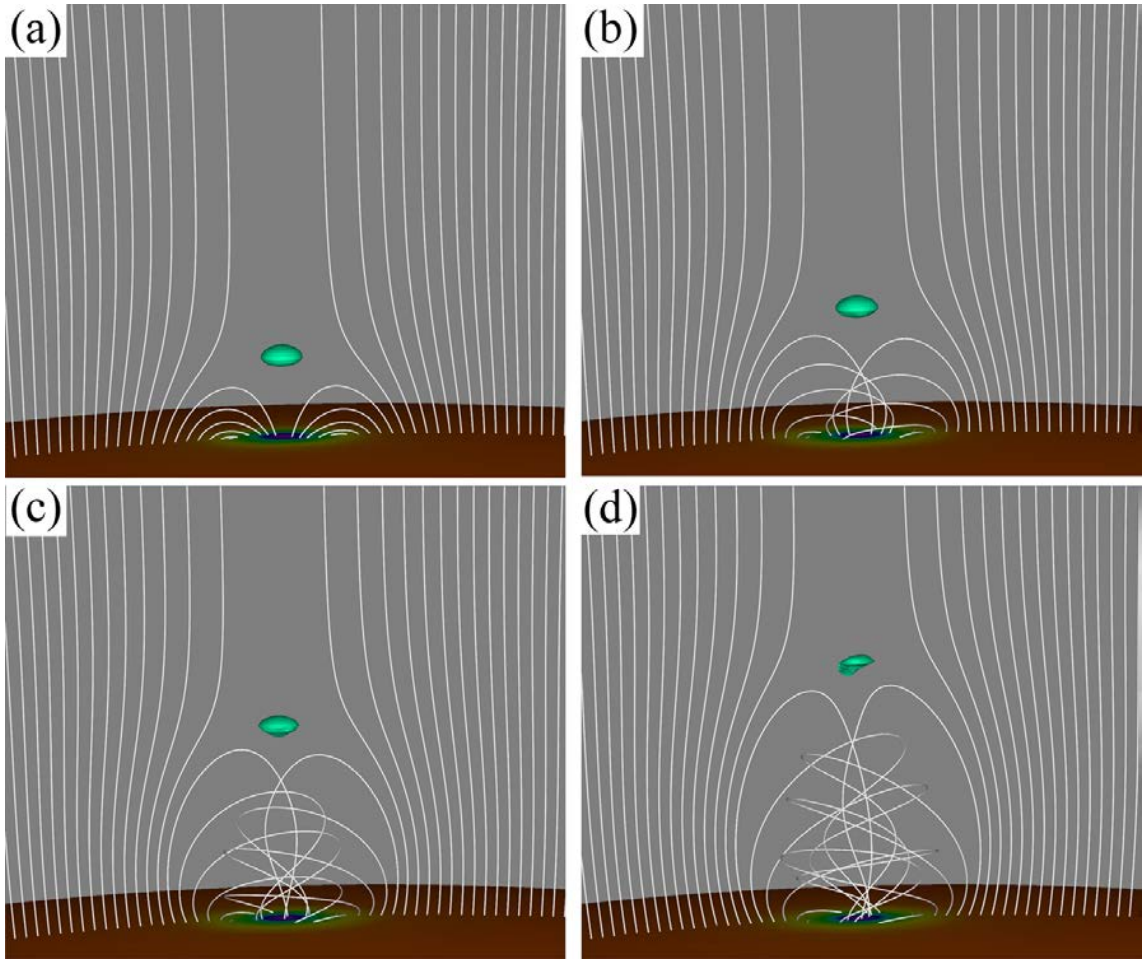

**Figure 4**. Close-in perspective view of the jet source region during the energy-buildup phase. White curves are magnetic field lines drawn from fixed footpoints at the lower boundary; color shading on the lower boundary indicates the sign and magnitude of the vertical magnetic field component as in Figure 1a; the green-colored spheroid is an isosurface of plasma $\beta = 2$ indicating the location of the null point and associated current patch. Snapshots are taken at times $t =$ (a) 625 s, (b) 1250 s, (c) 1875 s, and (d) 2525 s. Movie 1 is an animation of the full evolutionary sequence $t \in [0\ \text{s}, 4000\ \text{s}]$ at 25-s cadence.

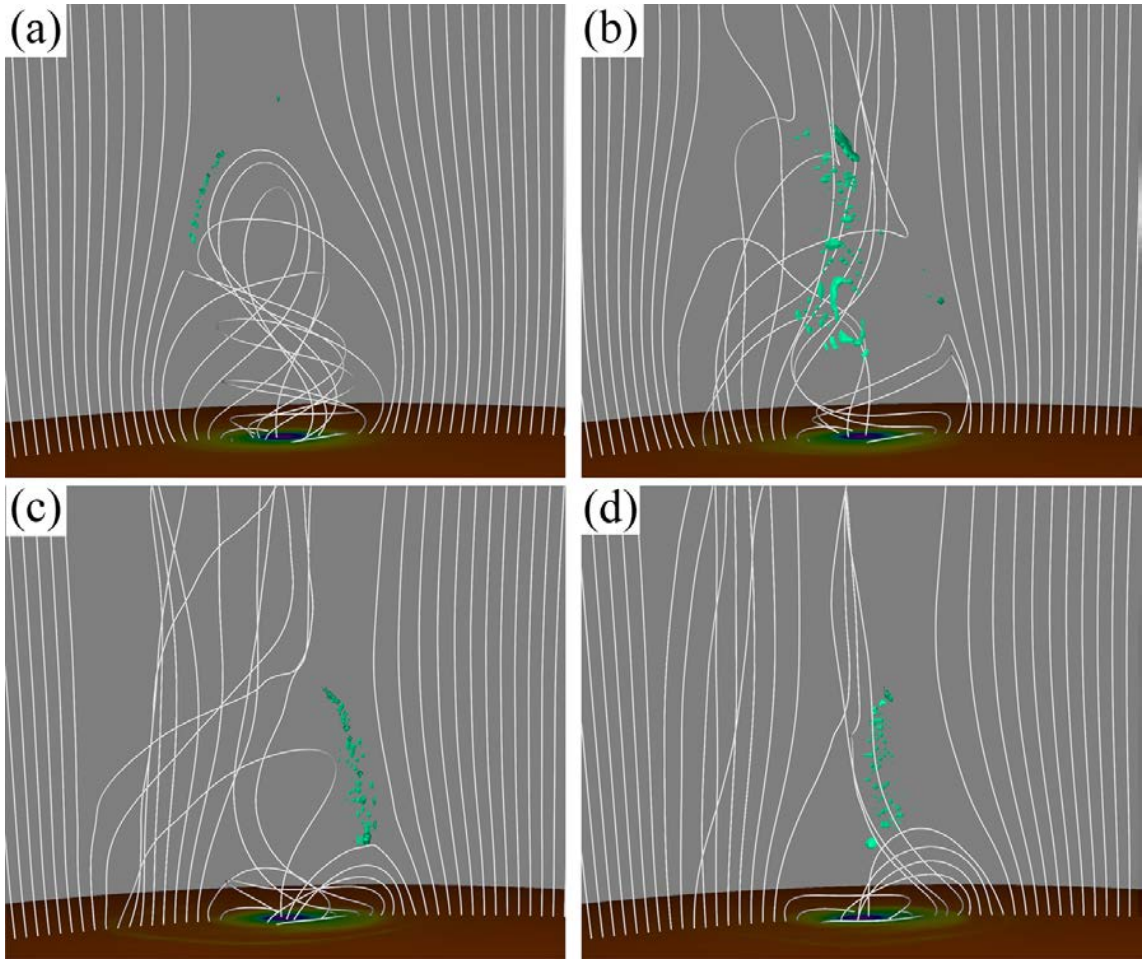

**Figure 5**. Close-in perspective view of the jet source region during the energy-release phase. Quantities displayed are the same as in Figure 4. Snapshots are taken at times *t* = (a) 2800 s, (b) 2925 s, (c) 3050 s, and (d) 3175 s. Movie 1 is an animation of the full evolutionary sequence *t* ∈ [0 s, 4000 s] at 25-s cadence.

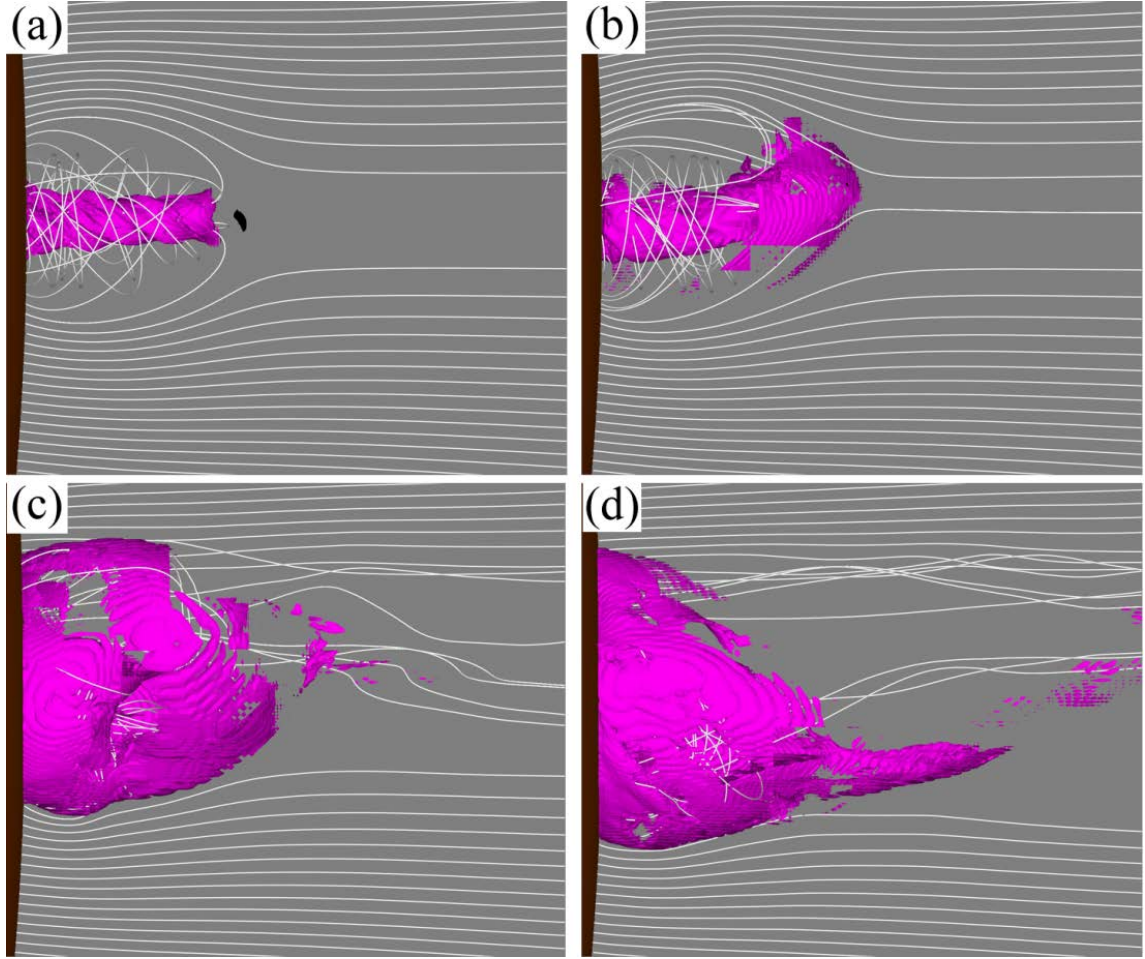

**Figure 6**. Close-in side view of the jet source region during the transition from energy-buildup to energy-release phases. White curves are magnetic field lines drawn from fixed footpoints at the lower boundary; the flattened black-colored volume is an isosurface of plasma $\beta = 2$; the magenta-colored region is an isosurface of current-density magnitude ($|\mathbf{J}|/c = 1.4\times10^{-9}$ G cm$^{-1}$); the coronal base is the dark surface to the left in each image. Snapshots are taken at times $t =$ (a) 2500 s, (b) 2750 s, (c) 2925 s, and (d) 3050 s. Movie 2 is an animation of the full evolutionary sequence $t \in$ [0 s, 4000 s] at 25-s cadence.

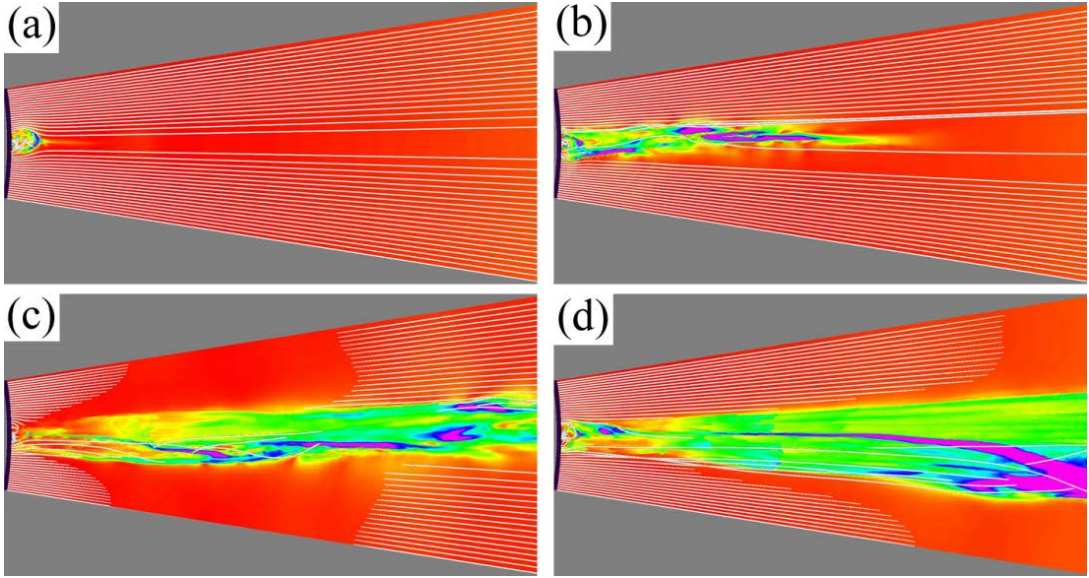

**Figure 7**. Mid-range view of the corona during the jet-propagation phase. White curves are magnetic field lines drawn from fixed footpoints at the lower boundary; color shading on the plane of the sky indicates the magnitude of the total plasma velocity, saturated at $5\times10^7$ cm s$^{-1}$; the coronal base is the dark surface to the left in each image. Snapshots are taken at times $t$ = (a) 2750 s, (b) 3150 s, (c) 3575 s, and (d) 4000 s. Movie 3 is an animation of the full evolutionary sequence $t \in$ [0 s, 4000 s] at 25-s cadence.

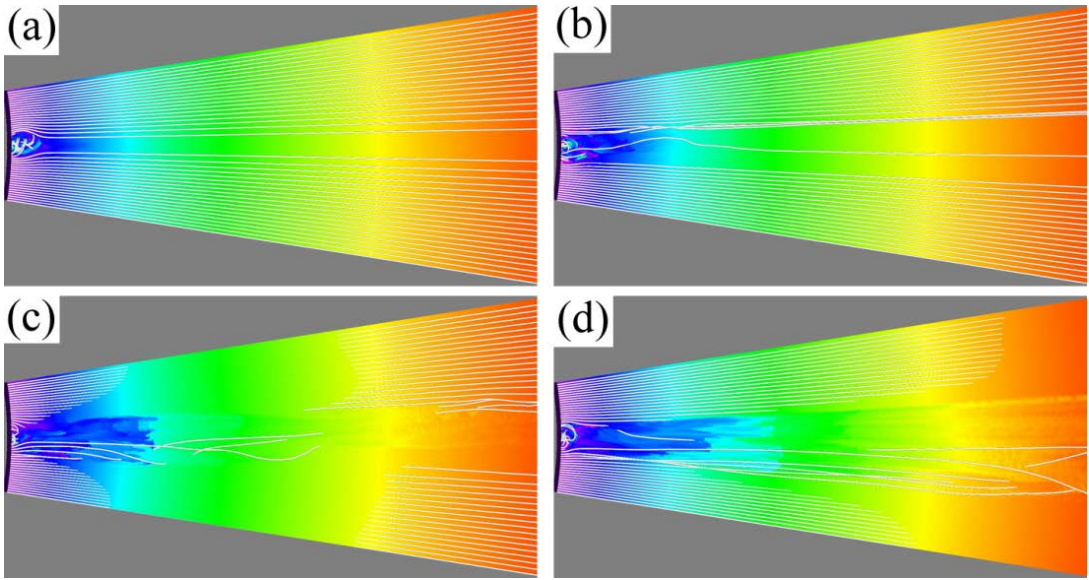

**Figure 8**. Mid-range view of the corona during the jet-propagation phase. Quantities displayed are the same as in Figure 7, except that color shading on the plane of the sky indicates the logarithm of the plasma mass density. Snapshots are taken at the same times as in Figure 7. Movie 4 is an animation of the full evolutionary sequence $t \in$ [0 s, 4000 s] at 25-s cadence.

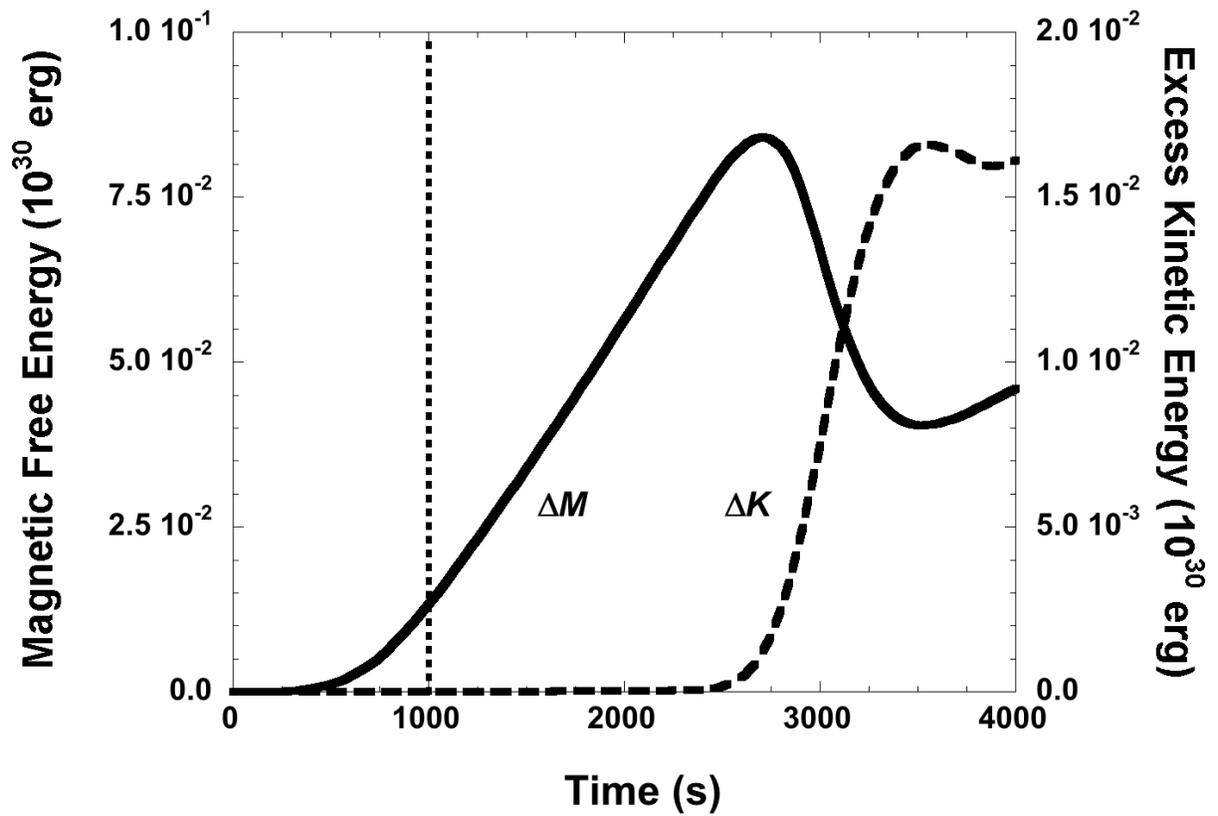

**Figure 9**. Magnetic free energy (Δ*M*; solid curve) and excess kinetic energy (Δ*K*; dashed curve) vs. time *t*, as defined in Equations (15) and (16). The vertical dotted line at $t = 1\times10^3$ s marks the transition from the ramp-up to the constant-driving phase.

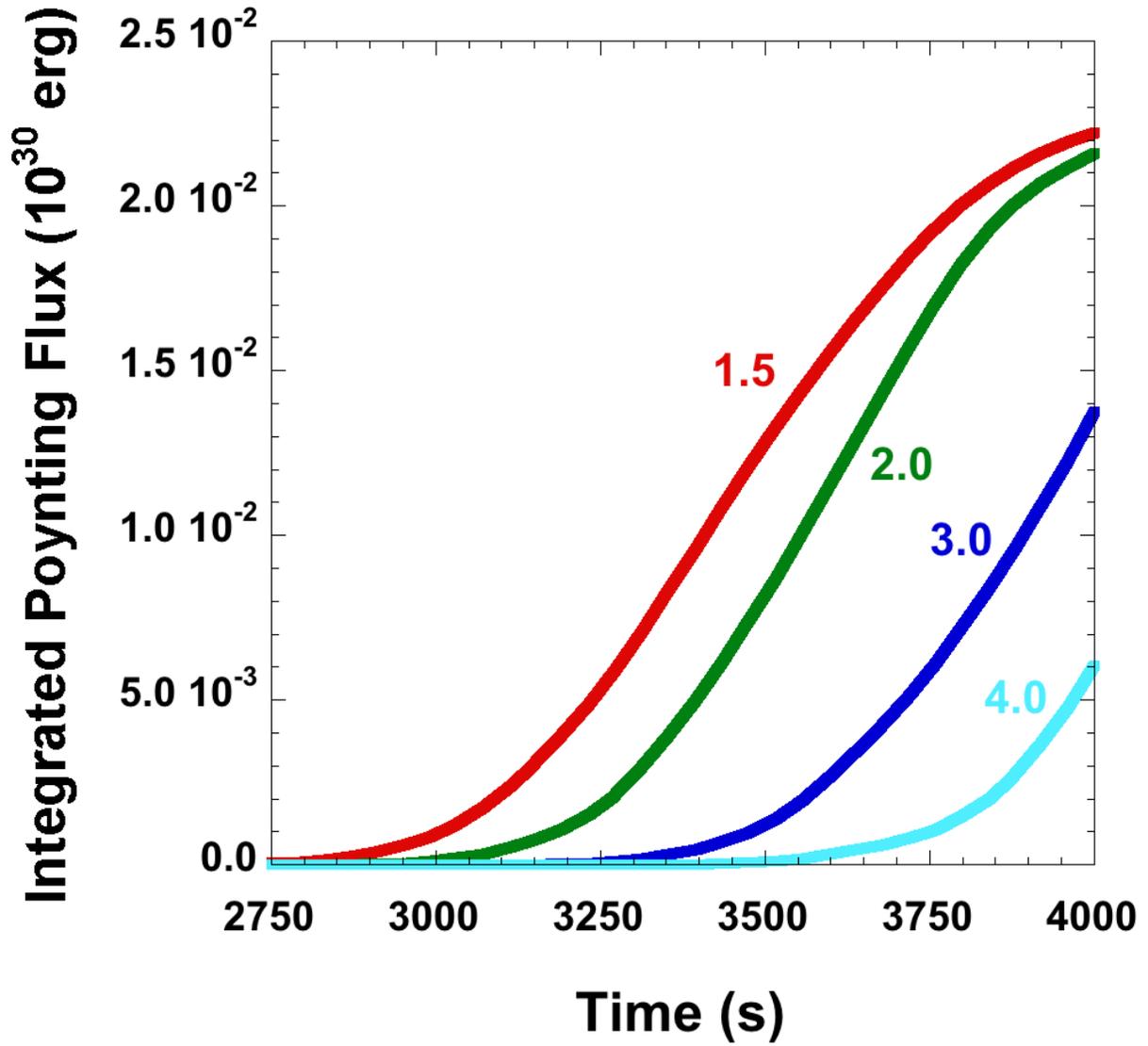

**Figure 10**. Time integrals of the Poynting fluxes of magnetic energy in Equation (17) evaluated at radii $r/R_\odot$ = 1.5 (red), 2.0 (green), 3.0 (blue), and 4.0 (cyan), as labeled.

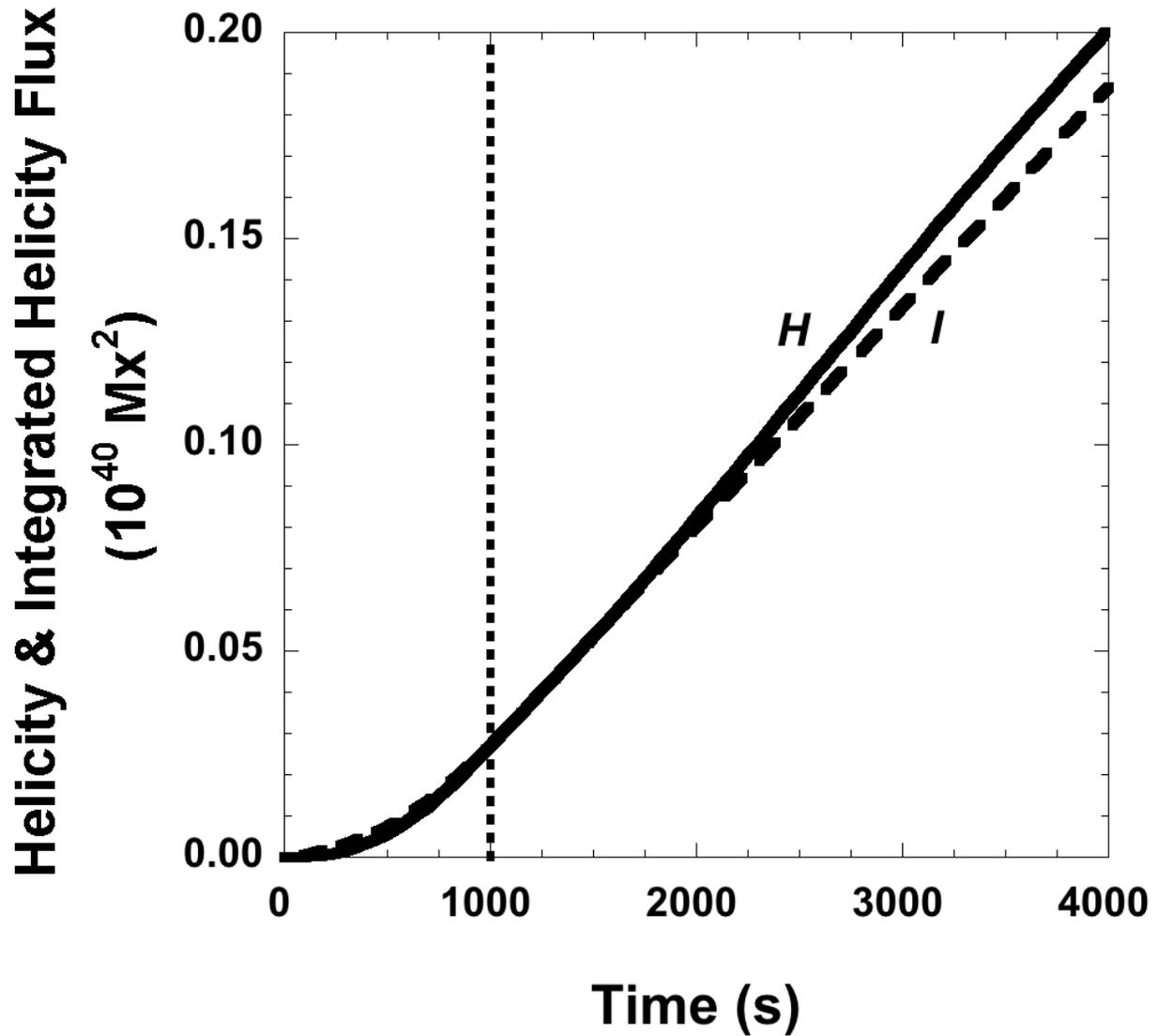

**Figure 11**. Volume helicity ($H$; solid curve) from Equation (19) and time integral of the helicity flux ($I$; dashed curve) from Equation (24). The vertical dotted line at $t = 1\times10^3$ s marks the change from ramp-up to constant-driving phases.

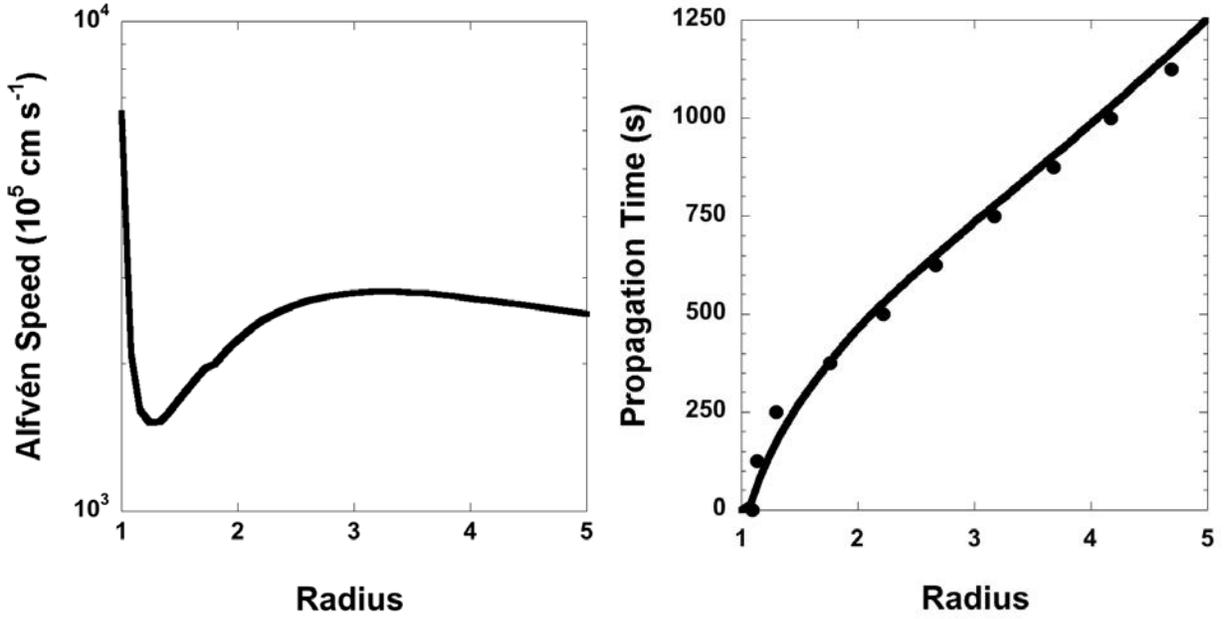

**Figure 12**. (a) Alfvén speed along the radial ray at $\theta = \pi/2$, $\phi = 0$, at the time of jet onset, $t = 2750$ s. (b) Propagation time of jet front, calculated from the local Alfvén and wind speeds according to Equation (28) (solid curve) and determined by finding the outermost extremum of $|\mathbf{B}_\nwarrow|$ (filled circles) at selected times.